\documentclass{article}
\usepackage[left=2cm, right=2cm, top=2cm, bottom=2cm]{geometry}
\usepackage{graphicx,amsmath,amssymb,psfrag,epsfig,multirow,color,comment,array,url,hyperref,setspace,pdflscape,rotating,marvosym}
\usepackage[mathlines]{lineno}
\usepackage{cite}
\usepackage{soul}
\usepackage{setspace}
\usepackage{bibentry}
\hypersetup{
	colorlinks,
	linkcolor={blue},
	citecolor={blue!50!black},
	urlcolor={blue!80!black}
}
\usepackage{amsthm}
\usepackage{amstext}
\usepackage{bbold}
\usepackage{comment}
\usepackage{booktabs}
\usepackage{csquotes}
\usepackage{enumerate}
\usepackage{epstopdf}
\usepackage{float}
\usepackage{physics}
\usepackage{rotating}
\usepackage{subcaption}
\usepackage{tabularx}
\usepackage{tikz}
\usetikzlibrary{shapes,arrows}
\usepackage[normalem]{ulem}

\usepackage{enumitem} 
\usepackage{kbordermatrix}
\usepackage{tabularx}

\usepackage[export]{adjustbox}
\usepackage[titletoc,toc,title]{appendix}

\newlist{todolist}{itemize}{2}
\setlist[todolist]{label=$\square$}
\usepackage{pifont}
\usepackage{color, colortbl}

\definecolor{LGray}{gray}{0.9}
\definecolor{DGray}{gray}{0.8}

\usepackage[aboveskip=1pt,labelfont=bf,labelsep=period,justification=raggedright,singlelinecheck=off]{caption}

\makeatother

\usepackage{authblk}
\hypersetup{
	colorlinks,
	linkcolor={blue},
	citecolor={blue!50!black},
	urlcolor={blue!80!black}
}
\newcommand{\red}[1]{\textcolor{black}{#1}}

\title{Age-specific transmission dynamics of SARS-CoV-2 during the first two years of the pandemic}

\author[1]{Otilia Boldea, PhD}
\author[1]{Amir Alipoor, MSc}
\author[2]{Sen Pei, PhD}
\author[2,3]{Jeffrey Shaman, PhD}
\author[4,5,6]{Ganna Rozhnova, PhD\footnote{\noindent{}Corresponding authors: \\
Dr. Ganna Rozhnova\\
Julius Center for Health Sciences and Primary Care \\
University Medical Center Utrecht \\
P.O. Box 85500 Utrecht, The Netherlands \\
Email: \href{mailto:g.rozhnova@umcutrecht.nl}{g.rozhnova@umcutrecht.nl}, phone: +31 631117965\\\\
Dr. Otilia Boldea\\Department of Econometrics and OR, and CentER\\
Tilburg School of Economics and Management, Tilburg University \\
P.O. Box 90153, Tilburg, The Netherlands\\
E-mail: \href{mailto:o.boldea@tilburguniversity.edu}{o.boldea@tilburguniversity.edu}, phone: +31 134663219}}
\affil[1]{Department of Econometrics and OR and CentER, Tilburg School of Economics and Management, Tilburg University, Tilburg, The Netherlands}
\affil[2]{Department of Environmental Health Sciences, Mailman School of Public Health, Columbia University, New York, USA}
\affil[3]{Columbia Climate School, Columbia University, New York, USA}
\affil[4]{Julius Center for Health Sciences and Primary Care, University Medical Center Utrecht, Utrecht University, Utrecht, The Netherlands} 
\affil[5]{BioISI---Biosystems \& Integrative Sciences Institute, Faculdade de Ci{\^e}ncias, Universidade de Lisboa, Lisbon, Portugal}
\affil[6]{Center for Complex Systems Studies (CCSS), Utrecht University, Utrecht, The Netherlands}

\usepackage{hyperref}

\begin{document}
\date{\today}
\maketitle
\newpage
\begin{abstract}
During its first two years, the SARS-CoV-2 pandemic manifested as multiple waves shaped by complex interactions between variants of concern, non-pharmaceutical interventions, and the immunological landscape of the population. Understanding how the age-specific epidemiology of SARS-CoV-2 has evolved throughout the pandemic is crucial for informing policy decisions. We developed an inference-based modelling approach to reconstruct the burden of true infections and hospital admissions in children, adolescents and adults over the seven waves of four variants (wild-type, Alpha, Delta, Omicron BA.1) during the first two years of the pandemic, using the Netherlands as the motivating example. We find that reported cases are a considerable underestimate and a generally poor predictor of true infection burden, especially because case reporting differs by age. The contribution of children and adolescents to total infection and hospitalization burden increased with successive variants and was largest during the Omicron BA.1 period. Before the Delta period, almost all infections were primary infections occurring in naive individuals. During the Delta and Omicron BA.1 periods, primary infections were common in children but relatively rare in adults who experienced either re-infections or breakthrough infections. Our approach can be used to understand age-specific epidemiology through successive waves in other countries where random community surveys uncovering true SARS-CoV-2 dynamics are absent but basic surveillance and statistics data are available. 

\end{abstract}

\section*{Introduction}

During the pandemic, the dynamics of SARS-CoV-2 demonstrated a complex spatio-temporal pattern with multiple waves \cite{WHO2022a} and pronounced differences in the age-specific burden of confirmed cases and hospitalizations \cite{COVID-19ForecastingTeam2022,ODriscoll2021}. A notable example is the age distribution of reported cases with much lower number of cases among younger individuals in European countries, including the Netherlands, reported in the first wave of the wild-type variant than in the spike of infections caused by the Omicron BA.1 variant of concern (VoC) \cite{coronadashboard}. 

Understanding how the age-specific epidemiology of SARS-CoV-2 has changed during the pandemic \cite{Koelle2022} is crucial for informing public health policy. For instance, information about 
how infection burden varies by age and time and the contribution of different age groups to transmission \cite{Monod2021,Davies2020} may inform non-pharmaceutical interventions \cite{Perra2021,Liu2021,Sharma2021,Brauner2021,Li2021} like school- and non-school-based measures \cite{Rozhnova2021}, while understanding of the age-specific hospitalization burden \cite{Flook2021} underpins prioritization of vaccination \cite{Bubar2021,Matrajt2021}. Some of this information can be provided by surveillance and serological surveys. However, full reconstruction of the age-specific burden of infections and hospitalizations in a country is complicated by several factors. Firstly, under-reporting of cases is age-specific and time-varying due to peculiarities of surveillance systems and testing policies, which varied across ages and time as the pandemic progressed. This fact coupled with the evidence of asymptomatic infection \cite{Sah2021} undermines the ability of surveillance to capture the true burden of infections among different age groups. Few countries conduct random community PCR testing \cite{ONS2022}. Secondly, nationally representative serological surveys provide information on which subpopulations carry antibodies to SARS-CoV-2 and thus could help to characterize prior infection burden \cite{Shioda2021,Hoze2021}. However, due to the waning of immunity after vaccination and infection, the immunological landscape of the population has become increasingly complex \cite{Milne2021}. Implementing representative serosurveys that estimate population immunity among different age groups characterized by varying numbers of prior infections before and after vaccination is difficult, costly, and time-consuming. Thirdly, additional factors including VoCs \cite{WHO2022b}, non-pharmaceutical interventions \cite{Perra2021,Liu2021,Sharma2021,Brauner2021,Li2021}, changes in population immunity after vaccination or infection, and seasonality in transmission \cite{Townsend2022,Kissler2020,Neher2020} complicate estimation of the age-specific burden.  

Modelling studies have provided important insights into temporal changes in the epidemiology of SARS-CoV-2 in specific countries \cite{Rozhnova2021,Viana2021,Pei2021,SaadRoy2020,Lavine2021,Kissler2020,Sonabend2021,Hoze2021}. Some of these were hypothesis-generating and not rigorously validated against all available evidence \cite{SaadRoy2020,Lavine2021,Kissler2020}. Other studies did not reconstruct  age-specific epidemiology \cite{Pei2021} or were limited to specific periods of the pandemic such as the first wave \cite{Rozhnova2021,Dekker2022,Gatto2020,Bertuzzo2020,Giordano2020} or periods of dominance for the Alpha \cite{Viana2021,Hoze2021} and Delta variants \cite{Sonabend2021,Bosetti2022}. As data accumulate, formal evaluations based on mathematical models calibrated to different types of observational data \cite{Pei2021,Rozhnova2021,Sonabend2021} are crucial for reconstructing the burden of infections and hospitalizations over long periods of time.

Here we reconstruct the epidemiology of SARS-CoV-2 in the Netherlands over the first 23 months of the pandemic, a period including wild-type, Alpha, Delta and Omicron BA.1 waves, using an age- and regionally stratified transmission model fitted to various data sources (see Supplementary Materials). Our fitting approach is a Bayesian evidence synthesis \cite{Presanis2011,Birrell2011,Rozhnova2020CMV,Rozhnova2021,Viana2021} based on an ensemble adjustment Kalman filter \cite{Anderson2001,Pei2020,Pei2021} combined with surveillance and national statistics data typically available from individual countries (hospital admissions, serological surveys, PCR testing data, genetic VoC data, vaccination coverage data, social contact matrices, demographic data, regional train and Google mobility data) with the Netherlands used as the motivational example. 

To account for control measures targeted at elementary and secondary schoolchildren versus the rest of the population \cite{Rozhnova2021} and for age-dependent transmission effects \cite{Mossong2008,Mistry2021,Goldstein2020,Viner2021,Sah2021,Flook2021,ODriscoll2021}, the population is stratified into young children (0 to 9 years old), adolescents (10 to 19 years old) and adults (above 19 years old) in twelve Dutch provinces. The model structure is rooted in current knowledge of SARS-CoV-2, which suggests waning of immunity after infection or vaccination \cite{Milne2021}, potential changes in susceptibility, infectivity and severity of re-infections and breakthrough infections \cite{Lavine2021,SaadRoy2020}, and seasonality in transmission \cite{Townsend2022,Kissler2020,Neher2020}. The regional stratification is augmented with real-world mobility across the provinces under the assumption that infected individuals with undocumented infection may travel to and infect susceptible individuals in other provinces whereas infected individuals with confirmed infection stay in their province of origin. Previous studies indicate that epidemic models with transmission dynamics coupled across locations can improve the identifiability of epidemiological parameters \cite{LiShaman2020,Pei2021}.

\section*{Results}
\subsection*{Time-dependent burden of hospital admissions}

\begin{figure}[!htbp]
\vspace{0.5cm}
\centering
\includegraphics[trim=0cm 0cm 0cm 0cm, clip=true, width=0.49\textwidth]{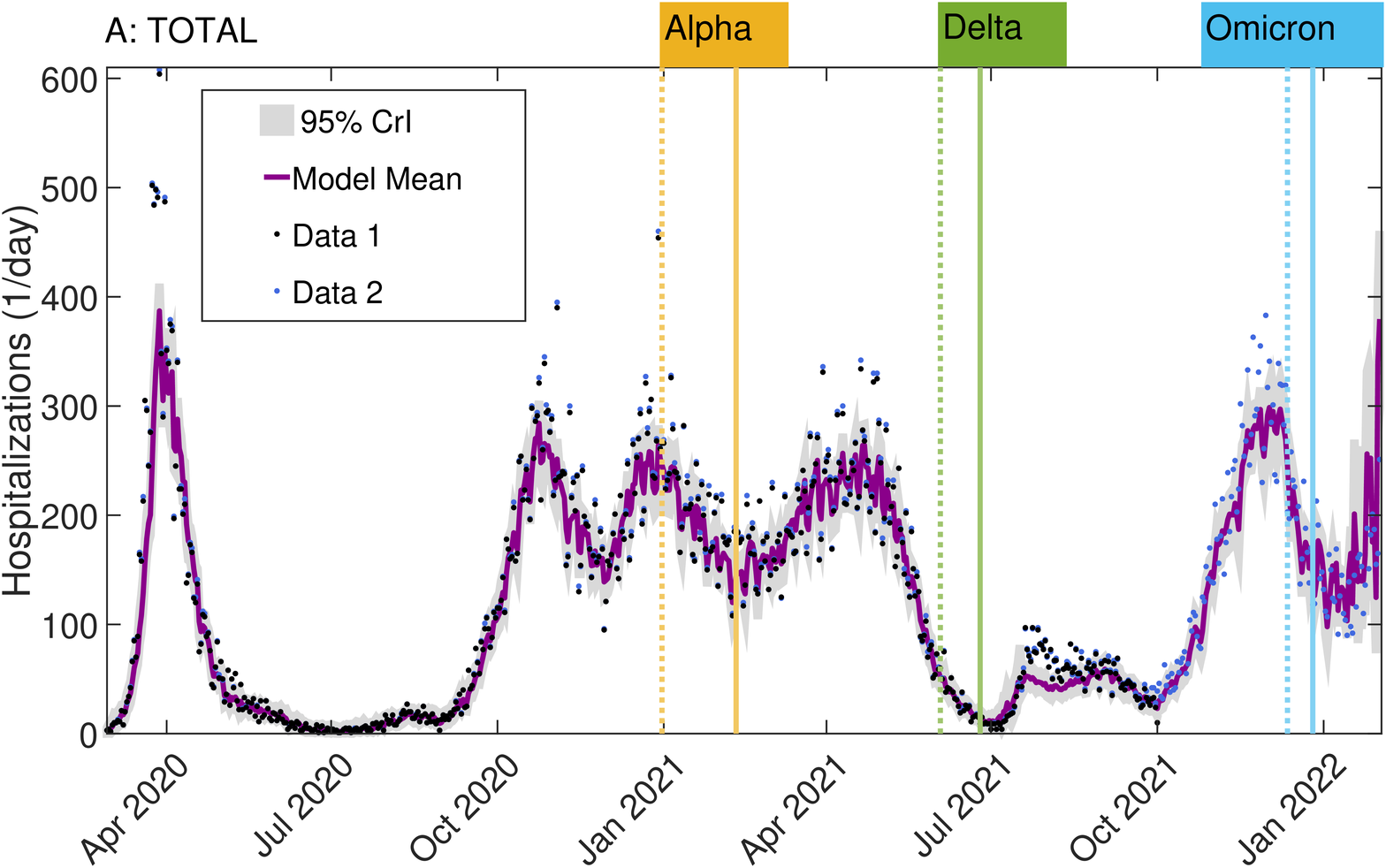}
\includegraphics[trim=0cm 0cm 0cm 0cm, clip=true, width=0.47\textwidth]{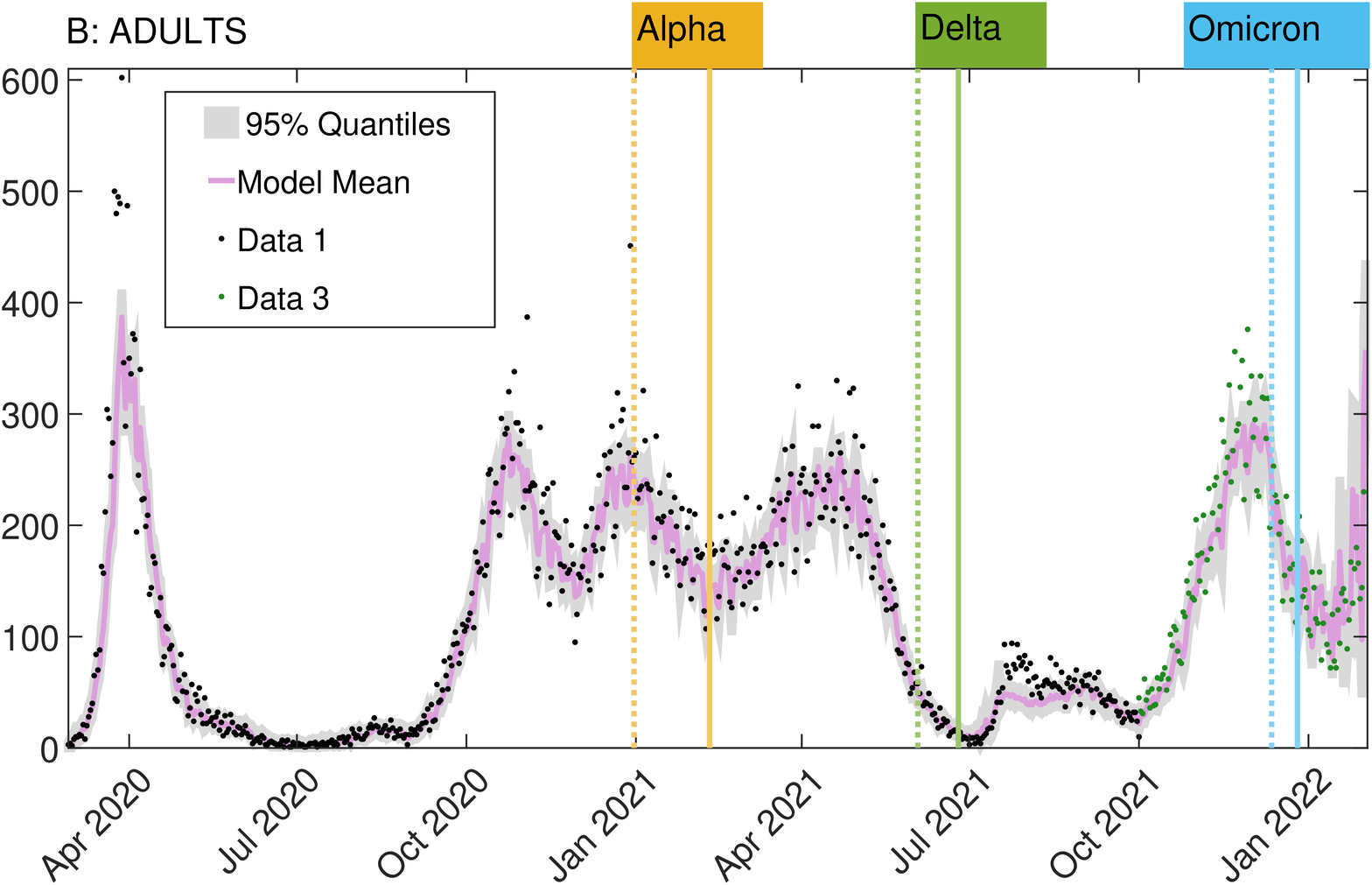}\\
\includegraphics[trim=0cm 0cm 0cm 0cm, clip=true, width=0.49\textwidth]{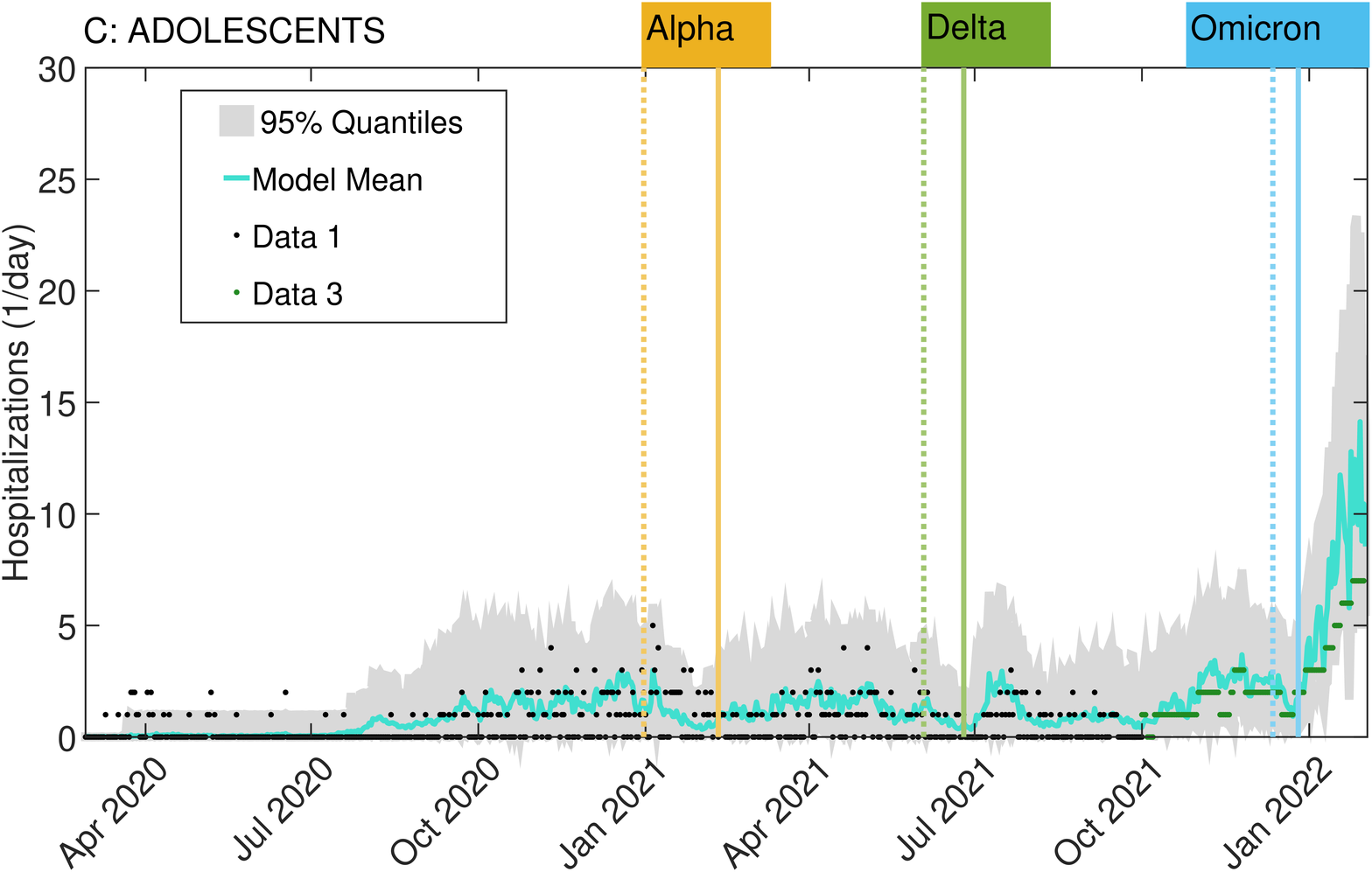}
\includegraphics[trim=0cm 0cm 0cm 0cm, clip=true, width=0.47\textwidth]{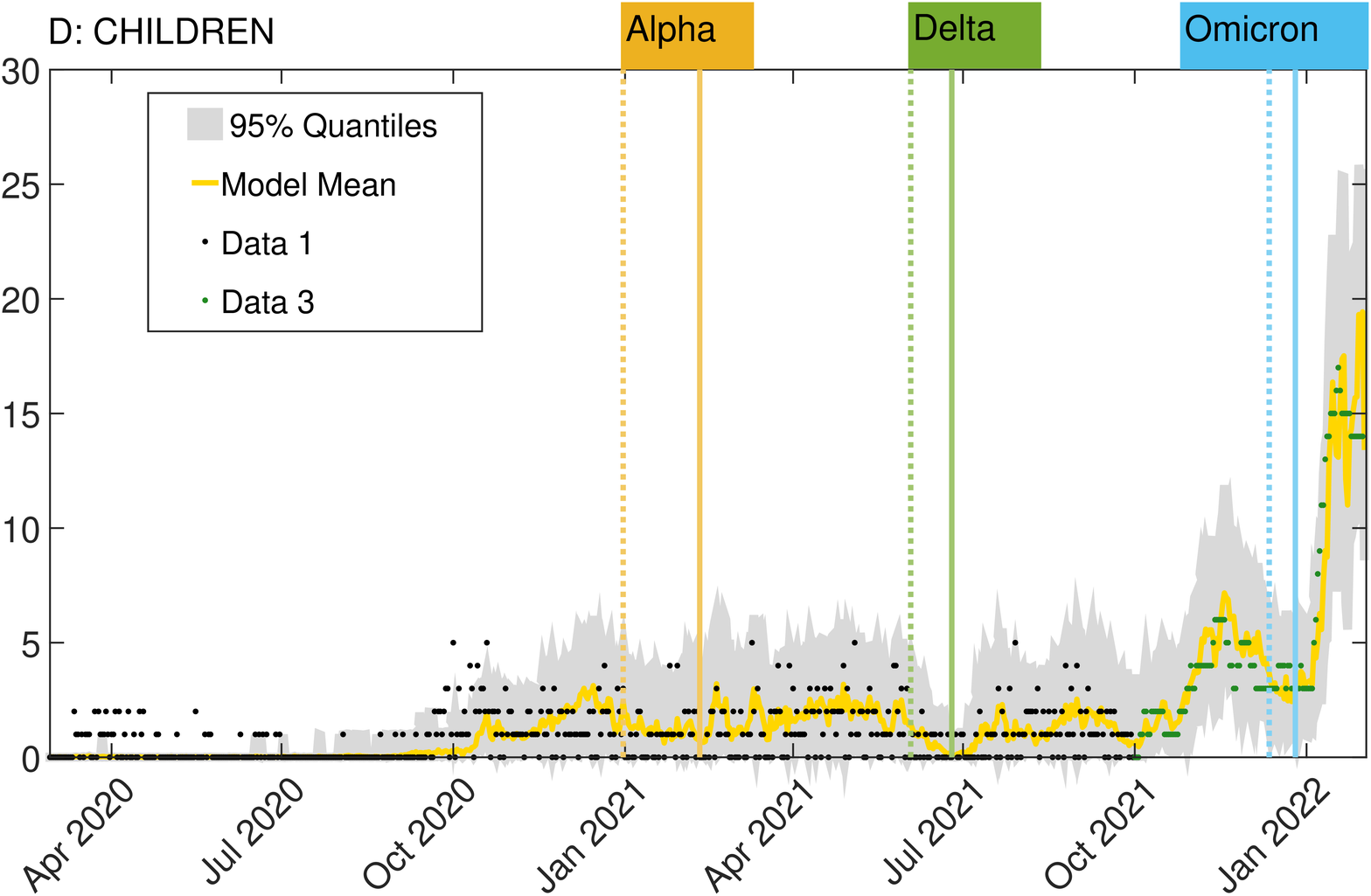}
\vspace{0.5cm}
\caption{{\bf Estimated hospital admissions.} Total (\textbf{A}) and age-specific national daily hospital admissions in adults (\textbf{B}), adolescents (\textbf{C}) and children (\textbf{D}). The colored lines represent the estimated posterior means for adults (pink), adolescents (turquoise), children (yellow), and all ages (purple). The gray-shaded regions correspond to posterior 95\% credible intervals defined as the 2.5\% and 97.5\% quantiles from 300 posterior ensemble values. The dots are daily hospital admission data used for fitting the model. \red{Data 1 refers to the National Intensive Care Evaluation data obtained from the National Institute for Public Health and the Environment (RIVM); Data 2 and 3 refer to the RIVM Dashboard data} (see Table \ref{tab1} and Supplementary Materials, Section 1). The dashed and solid vertical lines indicate when each VoC corresponded to 5\% and 50\% of samples in the \red{genetic variant} data, respectively.
}
\label{fig1}
\end{figure}

Our parsimonious model allows for robust estimation of key time-varying epidemiological parameters compatible with national and regional epidemiology of SARS-CoV-2 in pre- and post-vaccination periods (Supplementary Tables \red{A1---A6 and Figures A1---A4}; see also Section ``Parameter identifiability and sensitivity analyses''). The model reproduces well national and regional hospital admissions in young children, adolescents and adults during all waves that occurred from the first official case on February 27, 2020 until January 31, 2022 (Figure \ref{fig1}, Supplementary Figures \red{A5---A7}). At the national level, the total estimated mean number of hospital admissions is 84,674 (95\% CrI 82,729---86,151) (Supplementary Table \red{A7}) spread over seven waves in the wild-type, Alpha, Delta and Omicron BA.1 periods, with the last Omicron wave starting but not yet ending at the end of the study period, January 31, 2022 (Figure \ref{fig1} \textbf{A}). Of these, two and four waves occurred before and, respectively, after the start of vaccination, and one wave peaked in January 2021 around the onset of the vaccination program. As expected, the pattern of hospital admissions in adults largely mirrors that of total national admissions due to much higher probability of clinical disease and hospitalization in this subpopulation compared to adolescents and children \cite{Flook2021,ODriscoll2021,Rozhnova2021,Viana2021} (Figure \ref{fig1} \textbf{B}). In our analyses, we quantified the burden per period \red{of VoCs, defined as the time when a VoC reached a frequency of 5\% in the genetic variant data described in Table \ref{tab1}, until another VoC reaches the same frequency}. The burden of hospital admissions in adults progressively decreased over the periods of successive VoCs, in line with the expansion of the primary vaccination series and of the first booster campaign (Supplementary Table \red{A7}). The estimated cumulative mean number of hospital admissions \red{for adults} was thus largest for the wild-type (30,099, 95\% CrI 29,126---30,784) and smallest for the incomplete Omicron BA.1 period until January \red{31}, 2022 (8,352, 95\% CrI 7,895---8,816) (\red{Supplementary Table A7}). The picture is different for adolescents and children for whom the hospitalization burden stayed lower than 5 hospital admissions per day for either group and did not demonstrate a pronounced pattern until October 2021 (Figure \ref{fig1} \textbf{C} and \textbf{D}). Unlike in adults, the estimated cumulative mean number of hospital admissions in these two subpopulations was largest in the Omicron BA.1 and Delta periods (Supplementary Table \red{A7}; for adolescents, 292, 95\% CrI 239---346 and 266, 95\% CrI 173---453; for children, 433, 95\% CrI 333---586 and 388, 95\% CrI 258---693).

\subsection*{Time-dependent burden of confirmed cases and seroprevalence}

\begin{figure}[!htbp]
\vspace{0.5cm}
\centering
\includegraphics[trim=0cm 0cm 0cm 0cm, clip=false, width=0.47\textwidth]{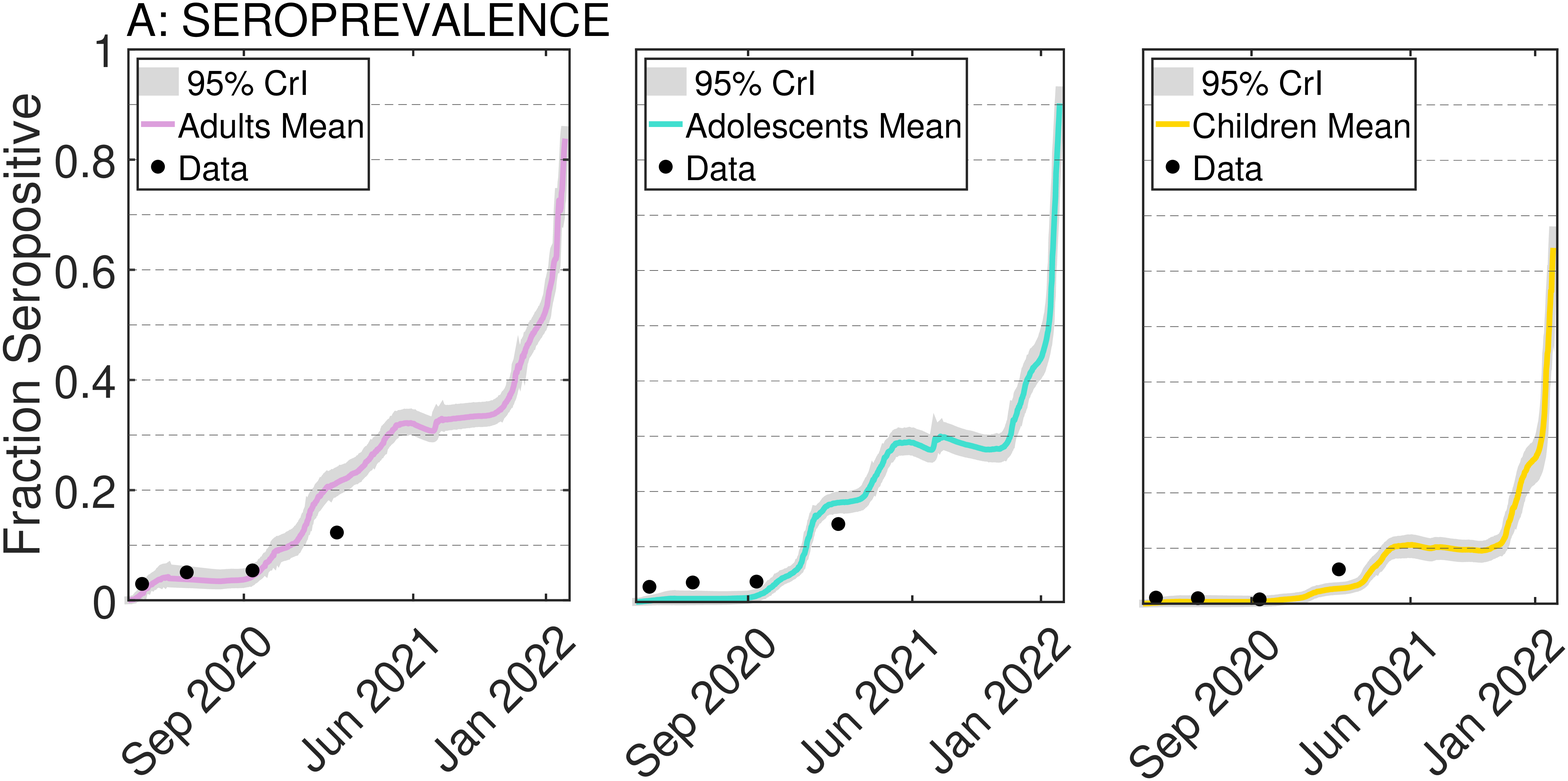}
\includegraphics[trim=0cm 0cm 0cm 0cm, clip=true, width=0.47\textwidth]{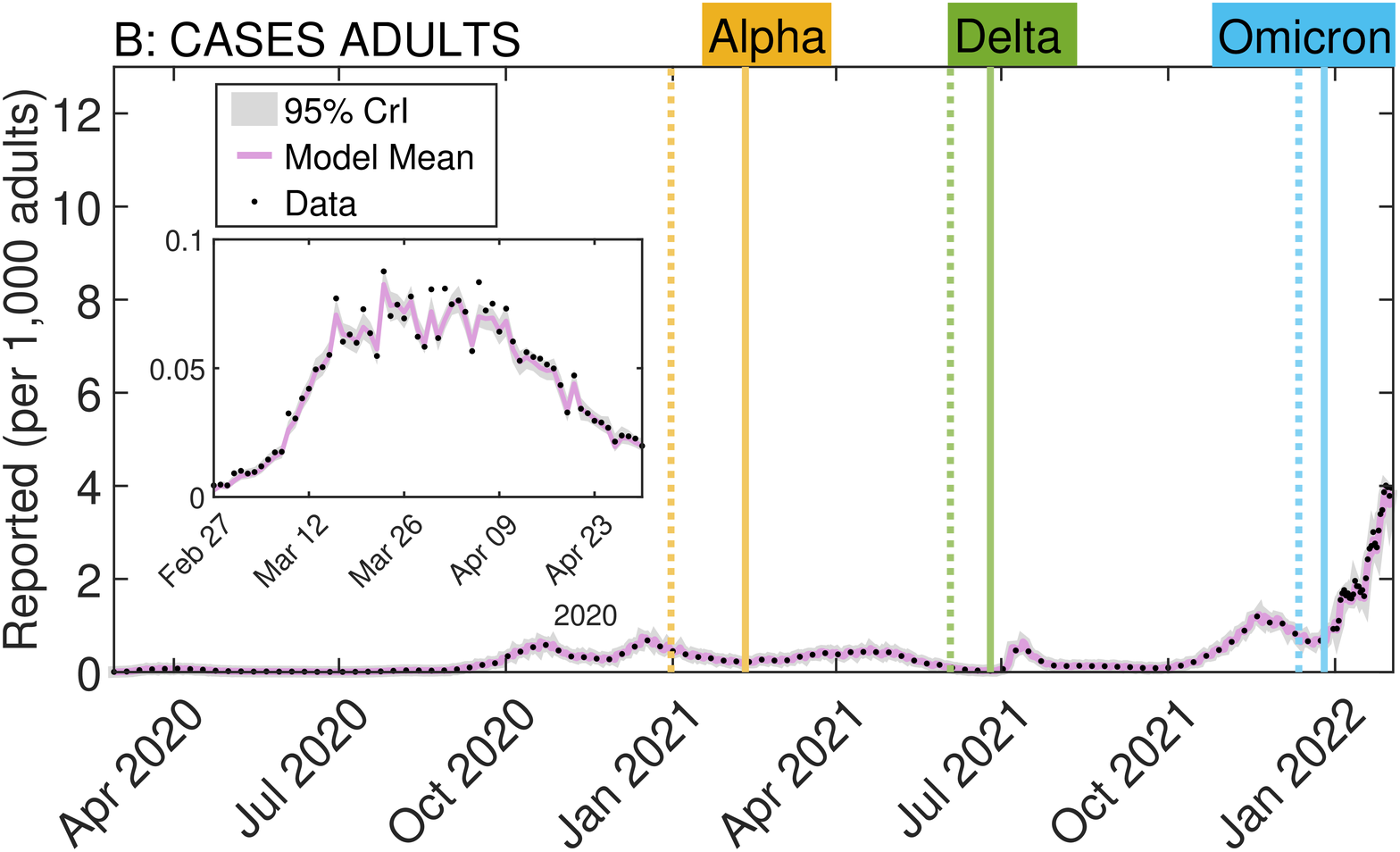}\\
\hspace*{0.1in}
\includegraphics[trim=0cm 0cm 0cm 0cm, clip=true, width=0.47\textwidth]{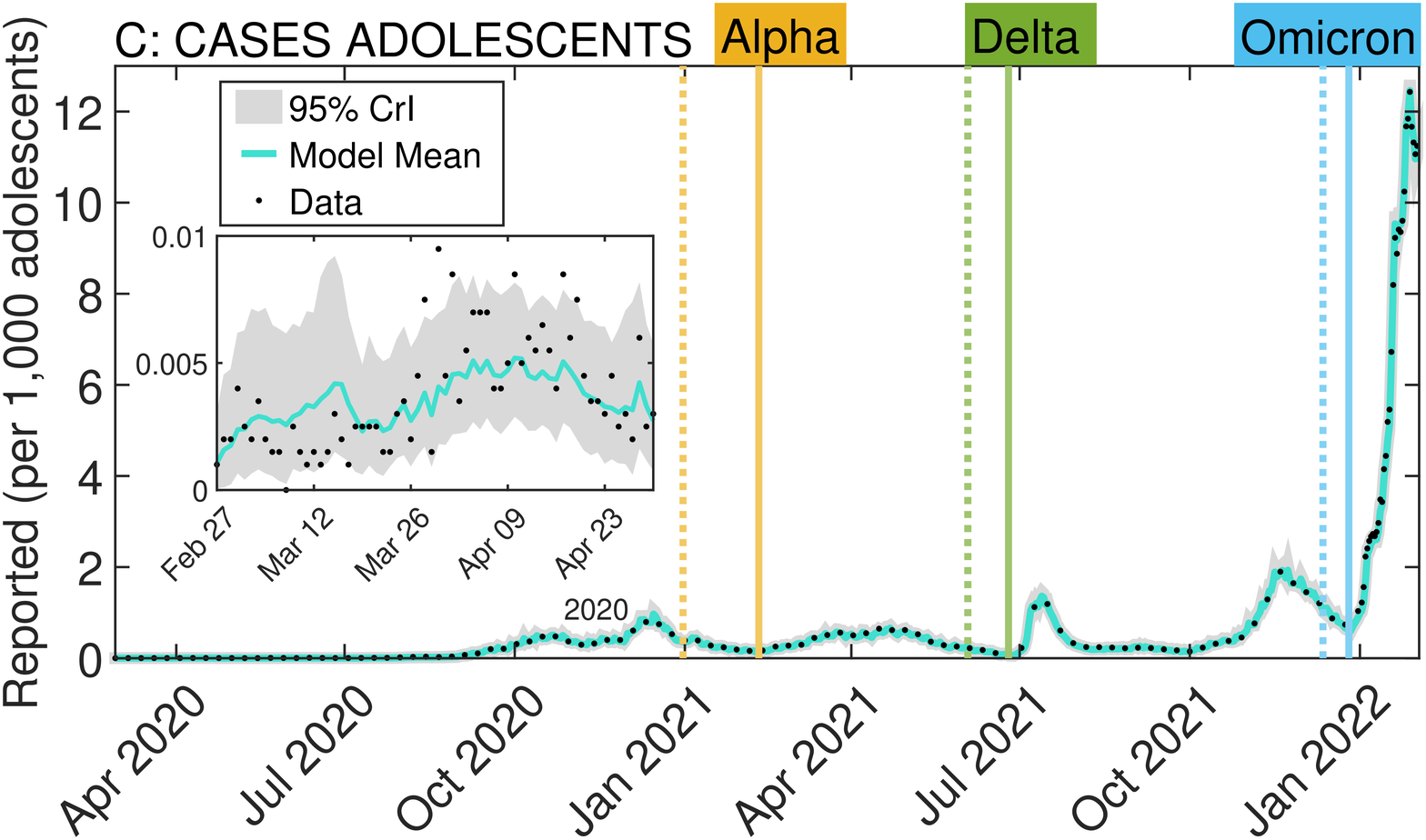}
\includegraphics[trim=0cm 0cm 0cm 0cm, clip=true, width=0.47\textwidth]{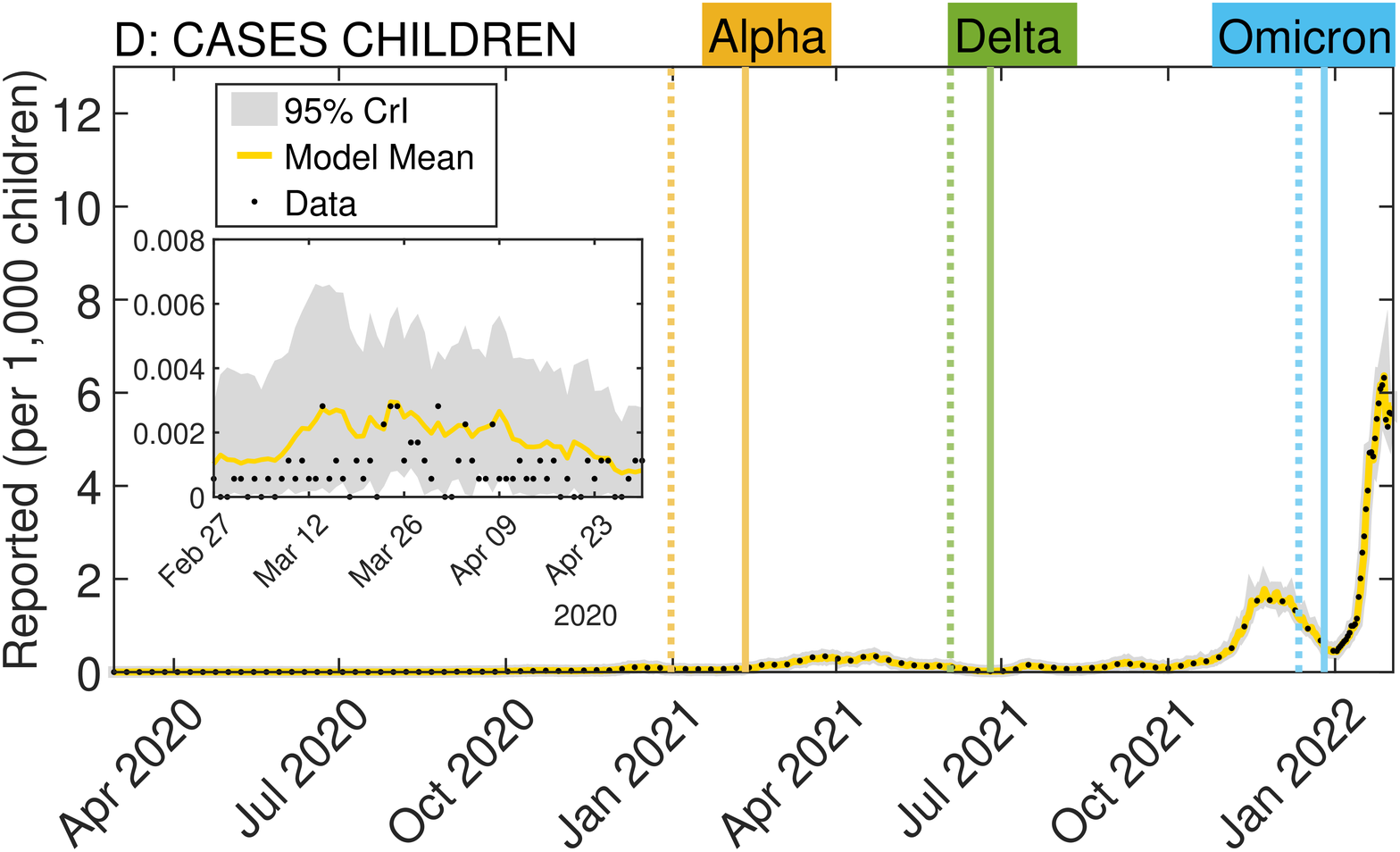}
\vspace{0.5cm}
\caption{{\bf Estimated seroprevalence and confirmed cases.} Age-specific national seroprevalence due to infection (\textbf{A}) and national confirmed daily cases in adults (\textbf{B}), adolescents (\textbf{C}) and children (\textbf{D}). Inset: the first wave characterized by low case reporting. The colored lines represent the estimated means for adults (\red{pink}), adolescents (turquoise) and children (yellow). The gray-shaded regions correspond to posterior 95\% credible intervals defined as the 2.5\% and 97.5\% quantiles from 300 posterior ensemble values. The black dots are seroprevalence (\textbf{A}) and daily confirmed cases (\textbf{B}, \textbf{C}, \textbf{D}) data used for fitting the model (see \red{Table \ref{tab1} and} Supplementary Materials, \red{Section 1}). The dashed and solid vertical lines indicate when each VoC corresponded to 5\% and 50\% of samples in the data, respectively. For comparison, the scale of the y-axis is the same in \textbf{B}, \textbf{C} and \textbf{D}.  
}
\label{fig2}
\end{figure}

\begin{figure}[!htbp]
\vspace{0.5cm}
\centering
\includegraphics[trim=0cm 0cm 0cm 0cm, clip=true, width=0.48\textwidth]{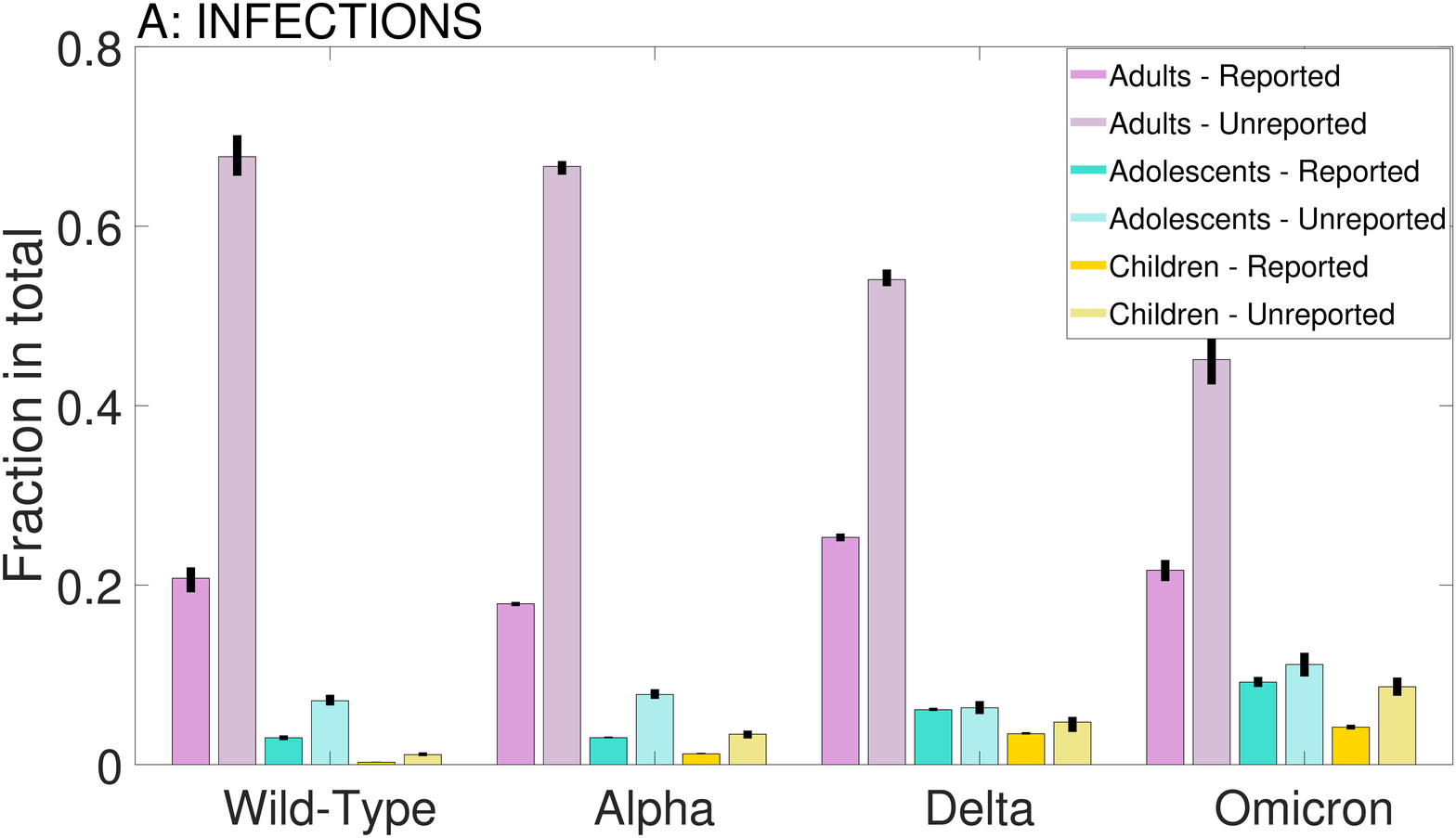}
\includegraphics[trim=0cm 0cm 0cm 0cm, clip=true, width=0.48\textwidth]{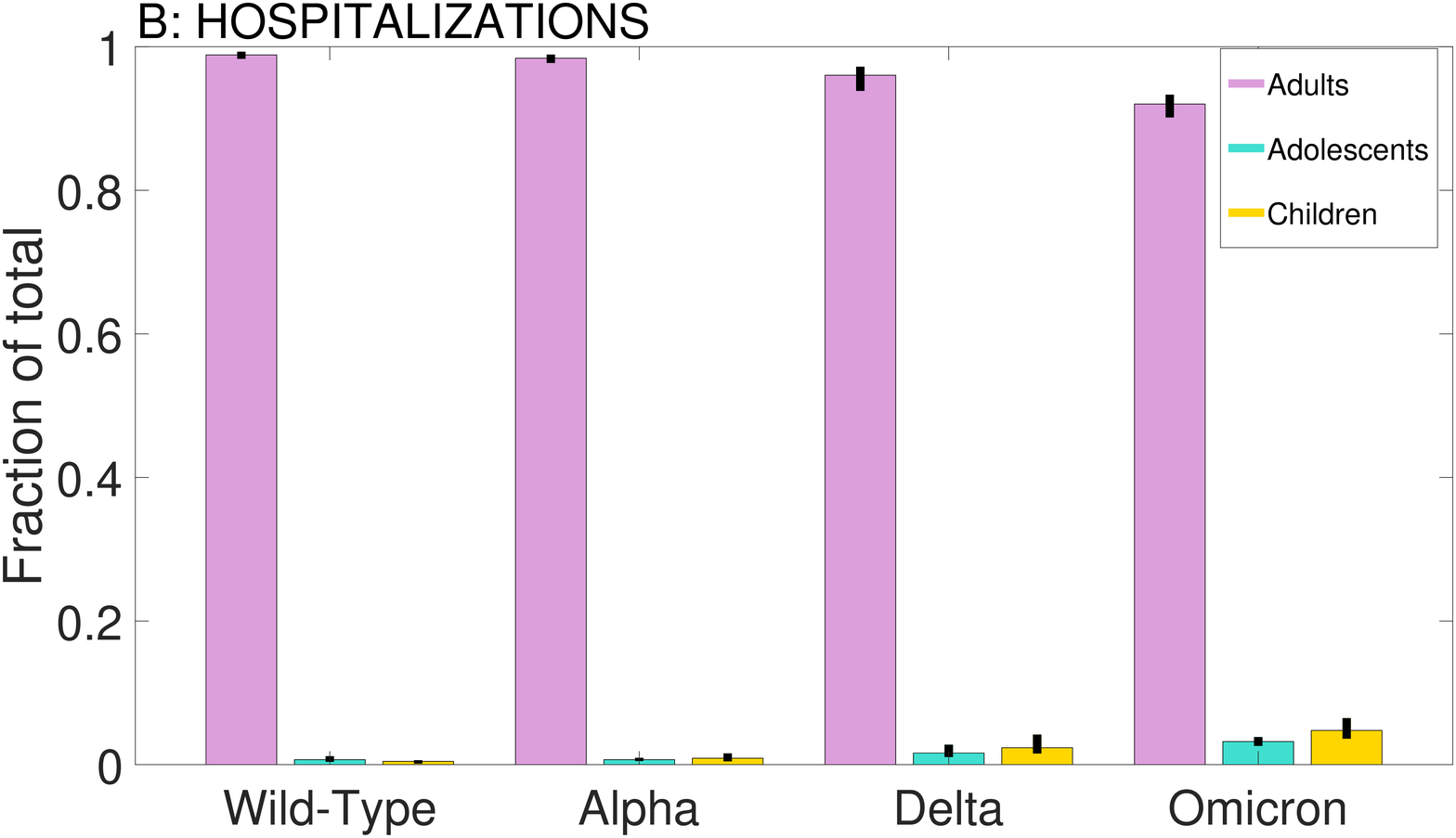}
\vspace{0.5cm}
\caption{{\bf Age distribution of estimated total infections and hospital admissions by VoC.} Age-specific fraction of confirmed and unconfirmed cases in the total national infections (\textbf{A}) and hospital admissions (\textbf{B}) during each VoC period. The bars and the black lines show the means and 95\% credible intervals obtained from 300 posterior ensemble values.
}
\label{fig3}
\end{figure}

The model further reproduces the age-specific seroprevalence and confirmed cases nationally \red{(Figure 2 \textbf{B}-\textbf{D})} and regionally (Figure \ref{fig2}, Supplementary Figures \red{A8---A16}). In the context of our study, the seroprevalence is the fraction of the population who have antibodies due to (any) infection. The level of seroprevalence is thus determined by how fast antibodies decay and how fast new infections happen. Confirmed cases refer to the part of infections captured by testing surveillance. The estimated national seroprevalence steadily increased in all age groups reaching 84\% (95\% CrI 82\%---86\%) in adults, 90\% (95\% CrI 87\%---93\%) in adolescents and 64\% (95\% CrI 60\%---68\%) in children by 31 January 2022 (Figure \ref{fig2} \textbf{A} \red{and Supplementary Table A8}). In contrast with the magnitude of the first wave of hospital admissions in spring 2020, the case reporting in this period was overall low, even more so among children and adolescents (insets Figure \ref{fig2} \textbf{B}, \textbf{C} and \textbf{D}). Throughout most of the study period that covers the next five waves, less than 1 case per 1,000 adults was reported daily, but \red{this} sharply increased during the Omicron BA.1 wave to 4 cases per 1,000 adults by the end of January 2022 (Figure \ref{fig2} \textbf{B}). The pattern of reported cases was similar in adolescents with less than 1 case per 1,000 adolescents daily during the first five waves; however, substantially more Omicron BA.1 cases were reported in this age group compared to adults during the sixth and seventh waves, skyrocketing from 1 to 12 daily cases per 1,000 adolescents in January 2022 (Figure \ref{fig2} \textbf{C}). More cases were documented in children than in adults in the Omicron BA.1 wave too, i.e., about 6 versus 4 cases per 1,000 individuals per day at the peak, while case reporting in children was much lower than in adults during the rest of the study period (Figure \ref{fig2} \textbf{D}). The observed increase in reported cases in adolescents and children compared to adults may indicate a higher burden of infections in these subpopulations during the Omicron BA.1 period and a change in the age-specific distribution of total national cases which we verify further below.

\subsection*{Age distribution of total infections and hospital admissions per VoC}

Our model estimates temporal changes in the age-specific contact rates as a result of implemented control measures (Supplementary Materials, \red{Section 2.4}). These changes combined with varying mobility patterns of reported and unreported cases across provinces allow estimation of the age-specific distribution of reported and unreported cases in the total national infections during each VoC period (Figure \ref{fig3} \textbf{A} \red{ and Supplementary Tables A9---A10}). The fraction of total infections in children among total estimated national infections (light and dark yellow bars) steadily increased from 1.4\% in the wild-type period to 12.9\% in the Omicron BA.1 period, of which about 66\% were unreported in total.
The same increasing contribution to total infections per successive VoC periods is estimated for adolescents (light and dark turquoise bars) who comprised between 10.1\% and 20.4\% of all national infections for the wild-type and Omicron BA.1 periods, respectively. 
The majority of all cases in adolescents were not reported to surveillance, with the estimated fraction of unreported infections 60\% in total for all VoCs. The contribution of adults to total infections (light and dark \red{pink} bars) decreased from 88.5\% to 66.8\% through the study period while the fraction of reported adult cases increased from approximately 23.5\% during the wild-type period to 32.5\% during the Omicron BA.1 period. In terms of the age-specific contribution to total hospital admissions per VoC, the adult population suffered the largest burden during all VoC periods with a minor fraction of hospital admissions attributed to children and adolescents during the circulation of Delta and Omicron BA.1 (Figure \ref{fig3} \textbf{B} and \red{Supplementary Table A11}).

\begin{figure}[!htbp]
\vspace{0.5cm}
\centering
\includegraphics[trim=0cm 0cm 0cm 0cm, clip=true, width=0.48\textwidth]{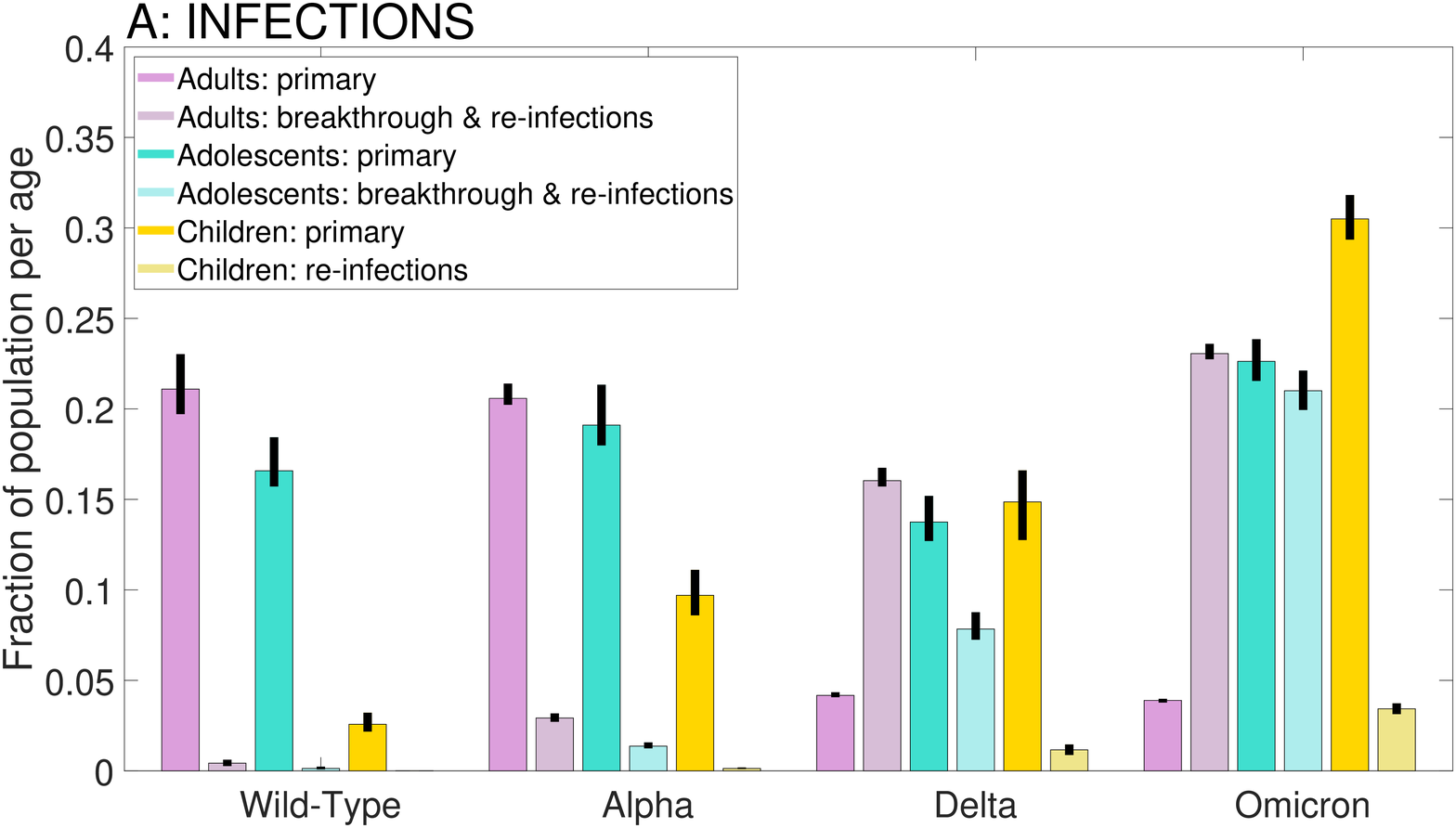}
\includegraphics[trim=0cm 0cm 0cm 0cm, clip=true, width=0.48\textwidth]{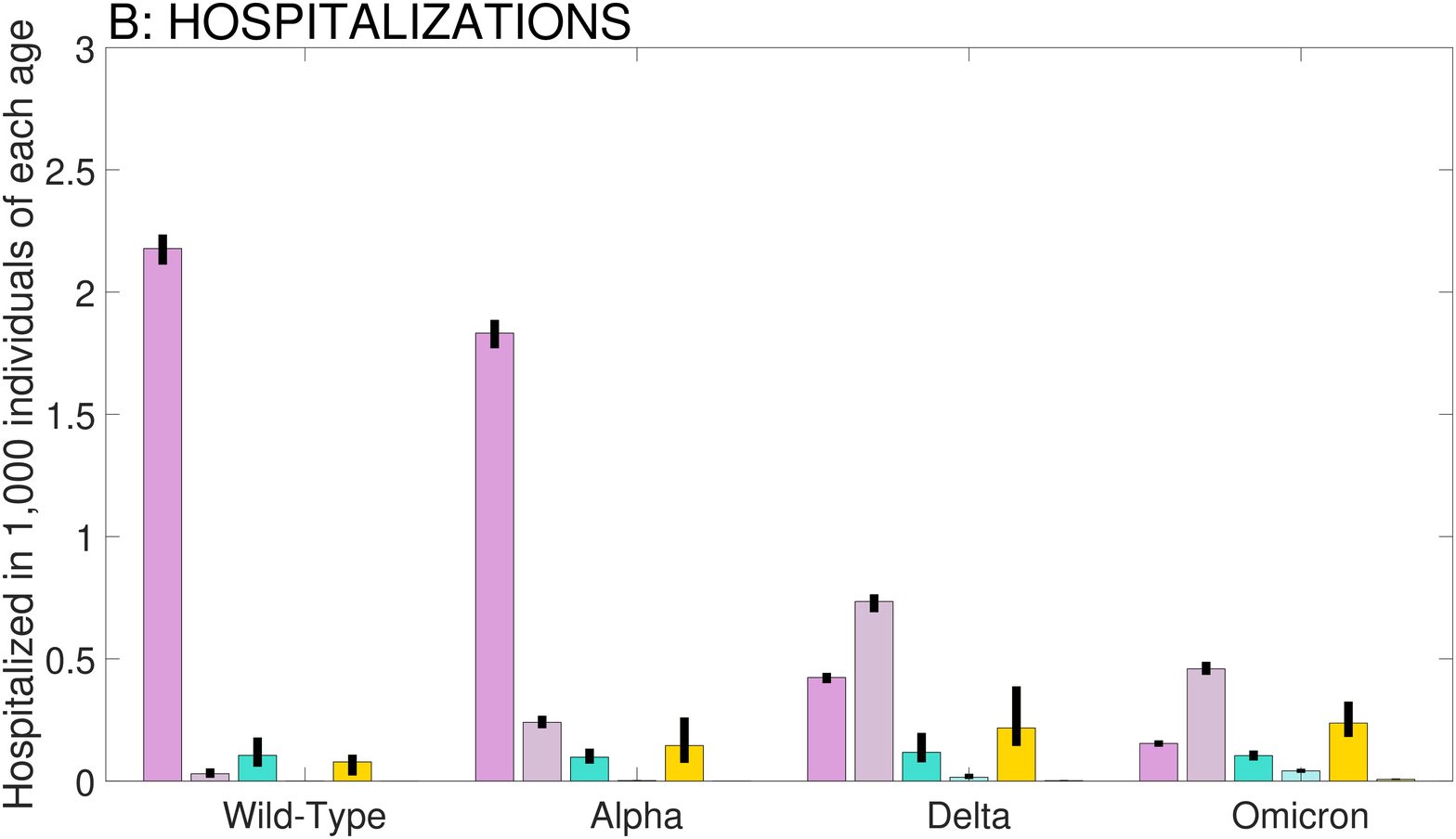}
\vspace{0.5cm}
\caption{{\bf Burden stratification by immune status and VoC.} (\textbf{A}) Fraction of primary and breakthrough infections or re-infections in the total population of each age group. (\textbf{B}) Hospital admissions after primary and breakthrough infections or re-infections per 1,000 individuals in each age group. The bars and the black lines show the means and 95\% credible intervals obtained from 300 posterior ensemble values.}
\label{fig4}
\end{figure}

\subsection*{Age-specific burden stratification by immune status and VoC}
We further analyzed the age-specific burden of total infections and hospital admissions stratified by immune status and period of VoC (Figure \ref{fig4} \red{and Supplementary Tables A12---A13}). For this, we distinguished infections in fully susceptible (naive) individuals (primary infections) from infections in individuals who were vaccinated (breakthrough infections) or lost immunity after primary infection (re-infections) (Figure \ref{fig4} \textbf{A}). Hospital admission burden was stratified into hospital admissions after primary infection versus after re-infection or breakthrough infection despite vaccination (Figure \ref{fig4} \textbf{B}). An estimated 21.1\% (95\% CrI 19.7\%---23.0\%) and 20.6\% (95\% CrI 20.2\%---21.4\%) of adults (\red{pink} bars) had primary infection during the wild-type and Alpha periods, respectively, while re-infections or breakthrough infections were experienced in only 3.3\% of adults for both periods combined. This pattern reversed during the Delta and Omicron periods as re-infections or breakthrough infections occurred in 16.0\% (95\% CrI 15.7\%---16.7\%) and 23.1\% (95\% CrI 22.7\%---23.6\%) of adults, respectively, while primary infections were experienced by only about 4\% of adults during each period. Similar dynamics are estimated for the burden of hospital admissions in adults; namely, there were more hospital admissions due to Delta and Omicron BA.1 breakthrough infections/re-infections than due to primary infections by these variants (Figure \ref{fig4} \textbf{B}). We stress that larger hospitalization burden in adults associated with breakthrough infections/re-infections than with primary infections during the Delta and Omicron BA.1 periods is a consequence of high vaccination coverage in adults while the hospitalization probability after vaccination or prior infection is very low compared to that in adults without any immunity. In the extreme case of a fully vaccinated adult population, all hospitalization burden would be attributed to breakthrough infections, and this would be much smaller than the burden the totally naive population experienced at the beginning of the pandemic. The burden of primary infections in adolescents (turquoise bars) was only slightly lower than that of adults for the wild-type (16.6\% of the age group, 95\% CrI 15.7\%---18.4\%) and Alpha (19.1\% of the age group, 95\% CrI 18.0\%---21.3\%) VoC. However, unlike for adults, primary infections in adolescents were more common than breakthrough infections or re-infections during the Delta and Omicron BA.1 VoCs, too (13.7\% and 22.6\% of the age group, respectively), owing possibly to the delayed vaccination schedule for adolescents. The burden in adolescents was especially high during the Omicron BA.1 wave when an estimated 43.6\% of all adolescents were infected. In part because young children were the last to start vaccination, the burden in this group increased with time and was mainly due to primary infections which, in turn, led to a slight increase of hospitalizations (yellow bars, Figure \ref{fig4} \textbf{A} and \textbf{B}). 

\subsection*{Parameter identifiability and sensitivity analyses}

\red{The posterior distribution of most parameters tighten after the first wave (Supplementary Materials, Section 6.1), indicating that the data are informative in recovering model parameters. To further verify system identifiability in the first wave, we generated one synthetic outbreak and verified that the generated infection and hospitalization data follow similar trends to the original data. To assess parameter identifiability in the first wave, we considered higher, lower and time-varying case detection rates. For each parameter combination, $100$ synthetic outbreaks were generated, each of which was used as data to re-estimate the model parameters. Across the three model configurations, the ``true" parameters were either within the range of the posterior mean densities, or the differences were not too large (Supplementary Materials, Section 6.2).}

Sensitivity analyses were conducted for the model parameters that were not estimated but rather calibrated (\red{see Methods and Supplementary Materials, Section 7}). The results were validated against Dutch seroprevalence estimates, \red{some of which} were not used in the estimation. The selected values for calibrated parameters yielded results that are closer to the Dutch seroprevalence estimates than alternative calibrations.

\section*{Discussion}

We developed an inference-based modelling approach to reconstruct how the age-specific epidemiology of SARS-CoV-2 changed through time and was shaped by the complex interaction between VoCs, non-pharmaceutical interventions, and the immunological landscape of the population. Our study provides several insights into changes in the age-specific burden of infections and hospitalizations over seven waves with different VoCs during the first two years of the pandemic in the Netherlands. Firstly, cases reported to surveillance were a considerable underestimate of total infections, as the majority of infections in all age groups remained unidentified despite overall improvement of case reporting through time. Reported cases are a generally poor predictor of true incidence of infection, especially because reporting differs substantially by age. Secondly, the contribution of children and adolescents to total infection and hospitalization burden increased with successive VoCs and was largest during the Omicron BA.1 period. This increase demonstrates that despite being recognized as less virulent than previous VoCs \cite{ECDC2022}, variants such as Omicron BA.1 can impose substantial burdens on adolescents and children and additional pressure on healthcare systems. Thirdly, we observed a shift in the pattern of infections and hospitalizations by immune status and age. Before Delta, almost all infections were primary infections occurring in naive individuals. However, during the Delta and Omicron BA.1 periods primary infections were common in children \red{who were infected less frequently early on and} for whom vaccination roll-out lagged \red{behind that of other age groups,} but were relatively rare in adults who experienced either re-infections or breakthrough infections. 

A similar shift in the dynamics was reported in a study based on identified case data from the UK \cite{Keeling2022}. However, unlike in \cite{Keeling2022} where re-infections dramatically increased during the Omicron wave, in our study the shift in primary infections versus re-infections and breakthrough infections is estimated to have begun during the Delta period. The discrepancy between the two studies could be attributed to the population stratification, used in this study. For computational efficiency, our model distinguishes fully naive individuals from those with immunity induced by either vaccine or natural infection, therefore re-infections and breakthrough infections are grouped.

To our knowledge, this is the first and most comprehensive application of a computationally efficient inference-based modeling approach that provides estimates of infection and hospitalization burden by age, region, VoC, and immune status \cite{Rozhnova2021,Miura2021,Ainslie2021,vanBoven2022,Uiterkamp2022,Gösgens2021,deVlas2021,Dekker2022}. The system is applied over seven waves during the first two pandemic years in the Netherlands. While we focus on the Dutch case, our framework can be applied to other countries given the common surveillance and national statistics data typically available for most countries. Compared to other studies \cite{Pei2021,SaadRoy2020,Kissler2020,Brand2021}, we separately model contacts in elementary schools, secondary schools, and the rest of the population to enable more reliable disentanglement of the roles of children, adolescents, and adults in transmission. We estimate contact changes due to non-pharmaceutical interventions outside the school environment based on proxy mobility measures \cite{KISHORE2022} --- Google and train mobility at transit stations. From this proxy, we observe that most measures are taken up by the population ahead of official implementation. We also estimate the speed of behavioral changes and the number of age-specific contacts after each intervention \cite{Viana2021,Rozhnova2021,Boldea2022}. This is in contrast to other inference-based studies (e.g. \cite{Sonabend2021}) that assume that age-related mixing patterns remain constant over time. Retrospective analyses of the impact on transmission and burden in different age groups of vaccination and boosting and of non-pharmaceutical interventions targeted at elementary and secondary schoolchildren, or at the rest of the population, will be a focus of future work.

Our model is fitted to regional reported cases, regional hospital admissions, and national seroprevalence. \red{We find that the first two data sources, combined with modeling movement of unreported infections across regions using mobility data, are crucial for identifying unreported infections with our inference method. The national seroprevalence data played a secondary role in reconstructing transmission dynamics due to the low frequency of serosurveys (i.e., four surveys during 23 months in our case) and the fact that our underlying transmission model is stochastic, unlike other studies \cite{Rozhnova2021,Viana2021}. Our approach could therefore be applicable to countries where seroprevalence estimates are not available at a high enough frequency.} Incorporation of other data into our model, particularly, publicly available SARS-CoV-2 measurements in sewage water, can help to address another problem that is increasingly apparent, i.e., drastic changes in COVID-19 surveillance in many European countries including the Netherlands. This model extension could provide projections of transmission dynamics after large-scale PCR testing stopped. Our model could also be extended to quantify SARS-CoV-2 infection and COVID-19 hospitalization burdens in vulnerable populations with pre-existing chronic conditions \cite{ECDC2021protocol} and to determine which interventions are required to protect them. 

Our model has limitations. Firstly, not testing children and adolescents in the early stages of the pandemic makes the identification of their age-specific parameters (i.e., susceptibility) weaker in that initial period. Later on, as testing capacity was expanded to younger individuals, we observe good identification of all age-specific parameters. Secondly, we did not model older age categories separately. It is technically straightforward to stratify the adult population into smaller age categories relevant for estimating the burden in older ages. However, older individuals travel less which means that our metapopulation model with mobility may not capture their case detection rates, and other data might be needed to estimate their age-specific parameters. Thirdly, our choice of stratification into primary versus breakthrough infections and re-infections is justified for the time period when a large proportion of the population did not yet have any immunity to SARS-CoV-2. As SARS-CoV-2 transitions from pandemicity to endemicity, primary infections will be experienced only by very young children born into the population \cite{Lavine2021,Cohen2022}. For later periods, our model could be modified to differentiate between several immunity classes such as individuals with primary vaccination series and various boosters, prior infections, and hybrid immunity \cite{Bobrovitz2022}.

In conclusion, we developed an inference-based transmission model that estimates how the age-specific epidemiology of SARS-CoV-2 changes over time. This approach is relevant for countries in which random community surveys uncovering true SARS-CoV-2 dynamics are absent but basic surveillance and statistics data are available. The findings of our study on the burden of infections and hospitalizations in children, adolescents and adults are important for informing public health policy on non-pharmaceutical interventions and vaccination.

\section*{Methods}
\subsection*{Overview}
The transmission model was calibrated using surveillance and national statistics data (PCR testing data, hospital admissions, serological surveys, demographic data, regional train and Google mobility data, vaccination coverage data, genetic VoC data, social contact matrices) for the Netherlands. Parameter estimates were obtained from the model fit to (i) age- and province-stratified SARS-CoV-2 case notification data during the period from February 27, 2020 until January 31, 2022; (ii) age- and province-stratified COVID-19 hospital admission data for the same period; and (iii) cross-sectional age-stratified national seroprevalence data from four serosurveys assessed on April 3, June 4 and September 20, 2020, and February 11, 2021 (median dates). Additional data for the model input were: (iv) population by age and province on January 1, 2020; (v) daily commuters across twelve provinces computed from Dutch national train data during the period from February 1, 2020 until September 30, 2022, and Google mobility data from February 5, 2020 until January 31, 2022; (vi) daily full vaccinations per province and age category during the period from January 31, 2021 until January 31, 2022; (vii) weekly boosters administered during the period from November 21, 2021 until January 31, 2022; (viii) weekly genetic VoC data during the period from December 1, 2020 until January 31, 2022; and (ix) school and non-school contact matrices from three surveys: before the pandemic, April 2020 and June 2020. All model analyses were performed in MATLAB, versions 2021A and 2022A. Data cleaning was performed in STATA SE v16. The code for fitting the model over the entire sample period runs in approximately 30 minutes on a Windows 10 Dell laptop with an Intel Core i5 processor and without parallelization.

\subsection*{Data}
Table \ref{tab1} gives an overview of the data, notation and sources. More details on how each dataset was constructed for use in the model fitting are given in the Supplementary Materials, Section 1.

\begin{table}[htp!]
\centering
    \caption{{\bf Overview of the data used in the model fitting.}}
\label{tab1}
\begin{tabularx}{\linewidth}{llX}
\toprule
    \rowcolor{LGray}
\textbf{Data} & \textbf{Notation}$^{*}$ & \textbf{Source}\\
\hline 
Population of age group $k$ & $N_{ik}$ & CBS, January 1, 2020 \\
in province $i$ &&\url{https://opendata.cbs.nl/statline/portal.html?_la=en&_catalog=CBS&tableId=37259eng&_theme=1135}\\ \hline
Daily reported cases & $Inew_{ik}^{obs}(t)$ & RIVM Dashboard \\
&&\url{https://data.rivm.nl/covid-19/COVID-19_casus_landelijk.csv}\\ \hline
Daily hospital admissions & & \\
Dataset 1 & $Hnew_{ik,(1)}^{obs}(t)$ & RIVM data, from February 27, 2020 until September 30, 2021 \\
Dataset 2 & $Hnew_{i,(2)}^{obs}(t)$ & RIVM Dashboard, from October 1, 2021 onward\\
&& \url{https://data.rivm.nl/covid-19/COVID-19_ziekenhuisopnames.csv}\\
&& \url{https://data.rivm.nl/covid-19/COVID-19_ziekenhuisopnames_tm_03102021.csv}\\
Dataset 3 & $Hnew_{k,(3)}^{obs}(t)$ & RIVM Dashboard, from October 1, 2021 onward\\
&& \url{https://data.rivm.nl/covid-19/COVID-19_ziekenhuis_ic_opnames_per_leeftijdsgroep.csv}\\ 
&&\url{https://data.rivm.nl/covid-19/COVID-19_ziekenhuis_ic_opnames_per_leeftijdsgroep_tm_03102021.csv}\\ \hline
Seroprevalence data & & obtained from RIVM PIENTER Corona Study \cite{Vos2020,Rozhnova2021} and\\
Serosurvey rounds & $h=1,2,3,4$ & \url{https://www.rivm.nl/en/pienter-corona-study/results}\\
Median inclusion time & $t_h^{ser}$&  \\
Fraction seropositive& $ser_{hk}$& \\
Sample size & $n_{hk}$ &\\
\hline
Train mobility  & $M_{ij}(t)$ & computed from the Dutch Railways (NS) data, \\
between provinces $i$ and $j$& & from February 27, 2020 until September 30, 2021;  extrapolated until January 31, 2022 using Google mobility data\\
\hline
Fraction commuters & $w_{ij}$ & CBS, year 2019 \\
from province $j$ to $i$ && \url{https://www.cbs.nl/nl-nl/cijfers/detail/83628NED}\\ \hline
Google mobility & $y_g$  & \url{https://www.google.com/covid19/mobility/}\\
&& from February 7, 2020 onward \\
\hline 
Daily vaccinations & $V_{ik}(t)$ & constructed from RIVM Dashboard \\
&&\url{https://data.rivm.nl/covid-19/COVID-19_vaccinatiegraad_per_gemeente_per_week_leeftijd.csv}  \\
& &and RIVM data \\
&&\url{https://www.rivm.nl/en/covid-19-vaccination/archive-covid-19-vaccination-figures-2021}\\
\hline
Booster transition function & $g_{B,k}(t), k=1,2$ & fitted to RIVM data \\
&&\url{https://www.rivm.nl/en/covid-19-vaccination/archive-covid-19-vaccination-figures-2022}\\
\hline
Variants transition function & $g_{\ell}(t), \ell \in \{\alpha,\delta, o\}$ & fitted to RIVM Dashboard data \\
&& \url{https://data.rivm.nl/covid-19/COVID-19_varianten.csv} \\
\hline
Contact matrices & & computed from \cite{Mistry2021,Backer:2021}  \\ Elementary school contacts & $c_{ikk^*,ES}(t)$ & and additional information on school closures and vacations\\
Secondary school contacts & $c_{ikk^*,SS}(t)$ & and additional information on school closures and vacations\\
 All-setting contacts & $c_{kk^*}^{(r)}$, $r=1,2,3$ & calculated from \cite{Backer:2021}, $r$ refers to survey rounds \\
 Non-school contacts & $c_{kk^*,NS}^{(r)}$, $r=1,2,3$ & calculated from \cite{Mistry2021,Backer:2021}, $r$ refers to survey rounds \\
    \hline
    \bottomrule
    \end{tabularx}
    \begin{flushleft}
    $^{*}$The indices $i,j=1,\ldots, 12$ denote the province of the Netherlands. The indices $k,k^*$ denote the age group, namely $k,k^*=1$ --- adults ($>19$ years old), $k,k^*=2$ --- adolescents ($10-19$ years old), and $k,k^*=3$ --- children ($0-9$ years old). The index $h=1,2,3,4$ denotes the serosurvey round. The index $l = \alpha, \delta, o$ denotes Alpha, Delta, and Omicron VoCs.
    \end{flushleft}
\end{table}

\subsection*{Transmission model}
\begin{figure}[!htbp]
\vspace{0.5cm}
\centering
\includegraphics[trim=0cm 0cm 0cm 0cm, clip=true, width=\textwidth]{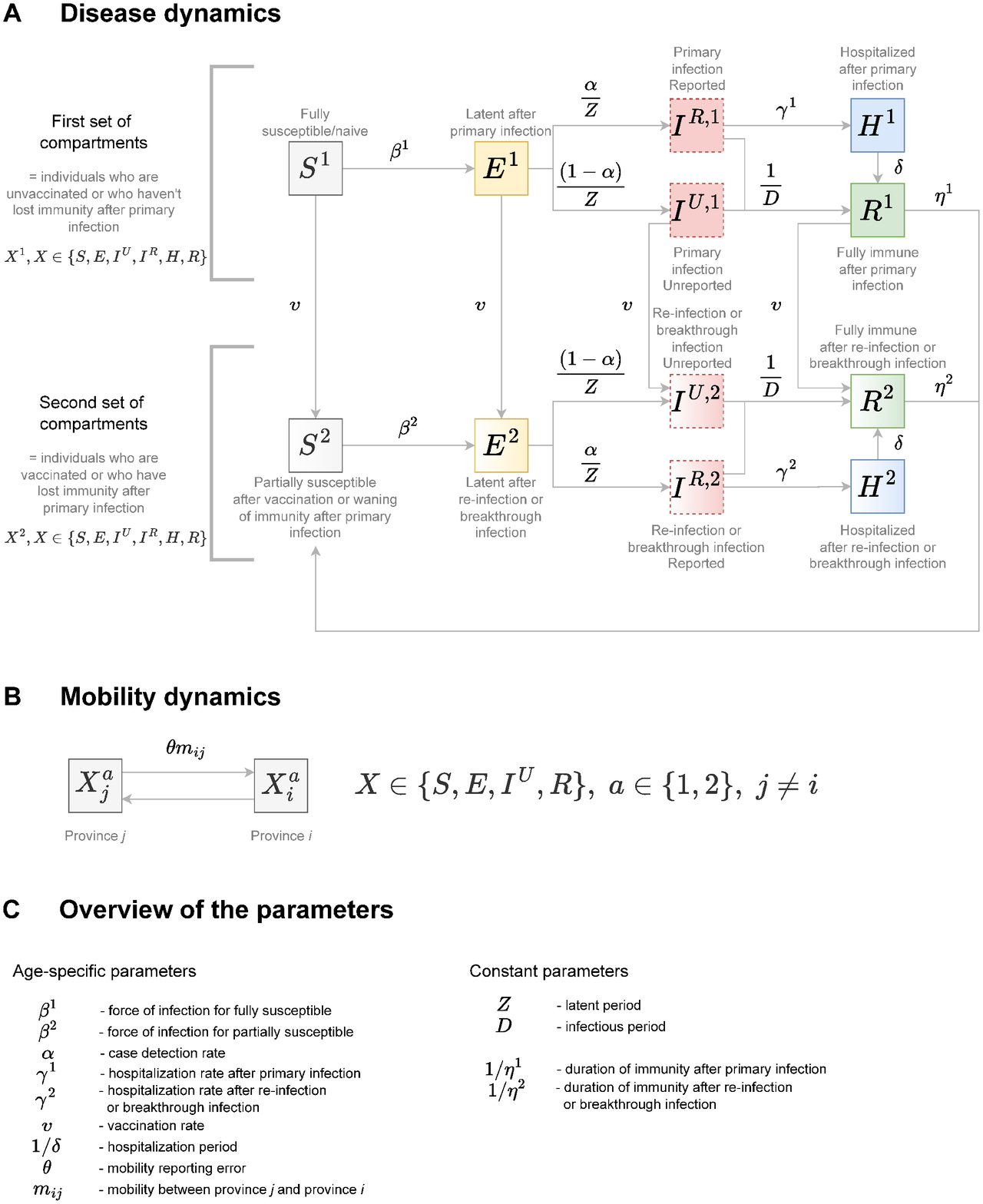}
\vspace{0.5cm}
\caption{{\bf Schematic of the metapopulation transmission model.} (\textbf{A}) Disease dynamics. (\textbf{B}) Mobility dynamics. (\textbf{C}) Overview of the parameters used in the model diagram. For simplicity of presentation, the age-specific indices are not shown in the schematic.}
\label{fig5}
\end{figure}

We developed a stochastic compartmental metapopulation model describing SARS-CoV-2 transmission in the population of the Netherlands stratified by province, disease status and age. A schematic depicting disease and mobility dynamics, as well as an overview of the main model parameters are shown in Figure \ref{fig5}.

\textbf{Disease dynamics} 
The fully naive susceptible individuals ($S^1$) in each age group can become latently infected ($E^1$) with the age-specific force of infection $\beta^1$. After an average latent period $Z$, the latently infected individuals become infectious (primary infection, $I^1$). Out of the total daily new infectious cases, an age-specific fraction $\alpha$ is reported to surveillance ($I^{R,1}$), and the $(1-\alpha)$ fraction is unreported ($I^{U,1}$). Both reported and unreported infectious individuals may recover without hospitalization ($R^1$) after an average infectious period $D$, but only reported infectious individuals may be hospitalized ($H^1$) with the age-specific rate $\gamma^1$. Hospitalized individuals are discharged after an average age-specific hospitalization period $1/\delta$. After a primary infection, individuals lose immunity and become partially susceptible ($S^2$) with the rate $\eta^1$. Disease progression for partially susceptible individuals is similar to that for fully susceptible individuals. However, partially susceptible individuals are reinfected with the age-specific force of infection $\beta^2$, individuals reported with re-infection are hospitalized with the age-specific rate $\gamma^2$, and the immunity after re-infection is lost with the rate $\eta^2$. Upon losing immunity after re-infection, the individuals return to the same partially susceptible compartment ($S^2$). Vaccination of individuals with the age-specific rate $v$ occurs in all disease stages except for those with primary reported infection or hospitalization. We used a simplified approach whereby susceptible individuals after vaccination and individuals recovered after primary infection whose immunity has waned are grouped together in one class of partially susceptible individuals ($S^2$). Therefore, re-infections and breakthrough infections are grouped, too. Vaccination or prior infection has three effects: (i) lower susceptibility to re-infection and breakthrough infection that affects $\beta^2$, (ii) lower infectivity of re-infection and breakthrough infection (not shown in the diagram), and (iii) lower hospitalization rate after re-infection and breakthrough infection $\gamma^2$. The model assumes a constant population size during the study period. 

\textbf{Mobility dynamics} 
The model assumes the unreported infected travel and infect other individuals in other regions, and that susceptible individuals can travel and become exposed in another region, while reported infected do not travel. Due to heterogeneity in population size and mobility across regions, this assumption equips the model with additional dynamics for the unreported infected compared to reported infected. The identification of unreported cases is possible as long as the case detection rate is the same across regions, and there are sufficient observations available across regions or time \cite{Li2020,Pei2021,Hortacsu:2021}. These two conditions are met in our study. Firstly, the case detection rate is assumed common across regions due to the ability to test at any location available across the country. Secondly, we estimate the model at a daily frequency for all twelve Dutch provinces.

A full description of the model and the model equations are reported in Supplementary Materials, Section 2.
\subsection*{Parameter inference}
The model is fitted to infection, hospitalization, and seroprevalence data using the ensemble adjustment Kalman filter \cite{Anderson2001}. This method allows inference on a high-dimensional system of observed and unobserved variables and parameters in a computationally efficient way. First, a large number of ensembles are drawn from priors on all parameters and state variables. The latter then evolve according to a stochastic version of the metapopulation model (Figure \ref{fig5}). For the filter updates, each ensemble member is individually propagated forward by means of a closed form approximation akin to the normal distribution but with an additional bias correction. The updates are computed sequentially based on the data in each province and its neighbours, allowing for a large number of state variables and parameters to be updated. Further details on the estimation and the algorithm can be found in Sections 3.3 and 3.4 of the Supplementary Materials. 

A small number of parameters that could not be identified were calibrated. These calibrations are described and motivated in Supplementary Materials, Section 3.1. The remaining parameters were estimated, and the choice of priors is detailed and motivated in the Supplementary Materials, Section 3.2. The parameter posteriors and their time-evolution over variant periods are discussed in Section 4 of the Supplementary Materials.

\subsection*{Model outcomes}
The exact calculations behind Figures \ref{fig1}---\ref{fig4} are explained in Supplementary Materials, Sections 5.1---5.4. The data plotted in Figures \ref{fig3} and \ref{fig4} can be found in Supplementary Tables A9 and A11---A13, respectively.
\subsection*{Details on the parameter identifiability and sensitivity analyses}
To verify system identifiability for the first wave, we fixed the parameters at their estimated posteriors in the original sample (observations from February 27, 2020 until March 30, 2020, see Supplementary Table A14), generated one synthetic outbreak and re-estimated the model parameters on the synthetic data. In addition to showing that the synthetic data are similar to the original data and are fitted well by our method, we also verified that the fitted and model seroprevalence match (Supplementary Materials, Section 6.1, Table A15 and Figures A17---A20). To assess parameter identifiability for the first wave, we considered three parameter configurations: (i) same as described above, with case detection rates equal to the original sample posterior means;  (ii) larger case detection rates; and (iii) case detection rates as in (i) until March 30, 2020, and as in (ii) until April 30, 2020. For each parameter combination, $100$ synthetic outbreaks were generated, each of which was used as data to re-estimate the model parameters. Across the three model configurations, the ``true" parameters were either within the range of the posterior mean densities or the differences were not too large (Supplementary Materials, Section 6.2, Figures A21---A23). In case (iii), the filter also approximated well the increase of case detection rates, even though this increase was not modelled. This suggests that changes in testing capacity or recommendations, and other unmodelled parameter changes can be captured by our model inference technique even in periods of high uncertainty, when data are less informative (Supplementary Figure A24).

For our sensitivity analyses, we re-estimated the model over the entire sample period: (i) without the serosurvey data; (ii) with lower and higher immunity waning rates after primary infection; (iii) with lower and higher immunity waning rates at the start of Omicron BA.1 period after both primary and breakthrough infection/re-infection; (iv) with higher transmissibility of Omicron BA.1; and (v) without seasonality. We found that in most cases, except (iii), our calibrations lead to closer matches to available seroprevalence estimates, independently of whether these were used in the model estimation or not (Supplementary Materials, Section 7 and Tables A16---A19). In cases (iii)-(iv), the results are robust to whether immunity wanes on average after two, three or four months for the Omicron BA.1 variant, and to whether Omicron BA.1 variant is 40\% or 60\% more transmissible compared to the Delta variant.



\section*{Reporting summary}
Further information on research design is available in the Nature Research Reporting Summary submitted together with this manuscript.

\section*{Data availability}
All output datasets generated during this study will be made publicly available in the designated GitHub repository (\url{https://github.com/oboldea/COVID_age_regions}) prior to acceptance of the manuscript. The datasets used as input in this study are listed in Table \ref{tab1}. All input datasets except for seroprevalence data and train mobility data will be made publicly available in the designated GitHub repository (\url{https://github.com/oboldea/COVID_age_regions}) prior to acceptance of the manuscript. Requests for seroprevalence data should be addressed to Fiona van der Klis (\href{fiona.van.der.klis@rivm.nl}{fiona.van.der.klis@rivm.nl}) and Eric Vos (\href{eric.vos.02@rivm.nl}{eric.vos.02@rivm.nl}) from the National Institute for Public Health and the Environment, Bilthoven, The Netherlands. Requests for train mobility data should be addressed to Jan Banninga (\href{jan.banninga@ns.nl}{jan.banninga@ns.nl}) from the Dutch Railways (Nederlandse Spoorwegen --- NS). 
All datasets are available to reviewers and editors upon request.

\section*{Code availability}
All codes reproducing the results of this study will be made
publicly available in the designated GitHub repository (\url{https://github.com/oboldea/COVID_age_regions}) prior to acceptance of the manuscript. The codes are available to reviewers and editors upon request.

\bibliography{nCoV}

\section*{Acknowledgements}
G.R. was supported by the VERDI project (101045989), funded by the European Union. Views and opinions expressed are however those of the author(s) only and do not necessarily reflect those of the European Union or the Health and Digital Executive Agency. Neither the European Union nor the granting authority can be held responsible for them. G.R. and S.P. were supported by the Fundação para a Ciência e a Tecnologia project 2022.01448.PTDC. O.B. and A.A. were supported by the Tilburg University Fund. J.S. and S.P. were supported by US NIAID grant R01AI163023 and CDC contract 75D30122C14289. We thank Prof. Marc Bonten (University Medical Center Utrecht), Dr. Susan van den Hof (National Institute for Public Health and the Environment), Prof. Dennis Huisman (Erasmus University Rotterdam) and Jan Banninga \red{(Nederlandse Spoorwegen) for their substantial help in obtaining the data for this study. We also thank Dr. Bettina Siflinger (Tilburg University) for the Tilburg CentER LISS Panel Survey questions added to support our study, Dr. Marino van Zelst (Wageningen University) and Prof. Jaap Abbring (Tilburg University) for useful discussions on data availability and modelling.}

\section*{Author contributions}
O.B. and G.R. developed the modeling study, with important feedback from all other authors. O.B. and G.R. obtained the data. O.B. and A.A. designed and wrote the computer code, with substantial feedback from S.P. G.R. wrote the manuscript, with substantial feedback from O.B., S.P., and J.S. O.B. wrote the Supplementary Materials, with substantial feedback from all other authors.

\section*{Competing interests}
J.S. and Columbia University declare partial ownership of SK Analytics. The other authors declare no competing interests.

\section*{Additional information}
Supplementary Materials contain details of the data, the model, inference, system identifiability, sensitivity analyses, Supplementary Figures, and Tables. Supplementary Materials are publicly available in the designated GitHub repository (\url{https://github.com/oboldea/COVID_age_regions}).

\section*{Correspondence}
\noindent{}Correspondence and material requests should be addressed to Dr. Ganna Rozhnova, Julius Center for Health Sciences and Primary Care, University Medical Center Utrecht, P.O. Box 85500 Utrecht, The Netherlands, email: \href{mailto:g.rozhnova@umcutrecht.nl}{g.rozhnova@umcutrecht.nl} and Dr. Otilia Boldea, Department of Econometrics and OR, and CentER, Tilburg School of Economics and Management, Tilburg University, P.O. Box 90153, Tilburg, The Netherlands, email: \href{mailto:o.boldea@tilburguniversity.edu}{o.boldea@tilburguniversity.edu}.

\end{document}